\documentclass{IEEEtran}
\usepackage{lipsum}
\usepackage{graphicx}

\title{Analysis and characterization of a twisted double loop antenna}

\author{
\IEEEauthorblockN{
    Danilo Paci\IEEEauthorrefmark{1}, 
    Clarbruno Vedruccio\IEEEauthorrefmark{2},
    Fabrizio Ciciulla\IEEEauthorrefmark{1}, 
    Dario Genovese\IEEEauthorrefmark{3}, 
    Gianni Albertini\IEEEauthorrefmark{1}\\
}
\IEEEauthorblockA{\IEEEauthorrefmark{1}Dipartimento SIMAU, Università Politecnica delle Marche, Italy\\}
\IEEEauthorblockA{\IEEEauthorrefmark{2}ICEMS, Milan, Italy\\}
\IEEEauthorblockA{\IEEEauthorrefmark{3}Dipartimento DIISM, Università Politecnica delle Marche, Italy\\}
}

\begin{document}

\twocolumn[
\begin{@twocolumnfalse}
\maketitle
\end{@twocolumnfalse}
\begin{abstract}
In this paper, we analyze the effects of twisting adouble loop torsional antenna. Its performance has been evaluated as a function of the twist angle around the symmetry axis from 0 to 225 degrees. Two geometric conditions are considered: in one case the height of the antenna is kept fixed. In the other, the height is varied so that the spherical shape is conserved. The results are compared with those of the corresponding finite element simulations. Increased Return Loss, increased bandwidth and decreasing frequency of resonance are obtained with respect to the untwisted case.
\end{abstract}
\bigskip]

\section{Introduction}
Changes of the resonance frequency and of the band-width are reported in literature when the geometry of an antenna is modified, without altering its size. Changes in center-band frequency and bandwidth are correlated with changes in the geometry of a patch antenna \cite{7843207} due to mechanical torsion of the support.\\
Variations in the bandwidth of the Electro Magnetic signal are also reported in a vibrating dipole \cite{7335594}.\\
Chaotic vibrations induced significant content beyond the fundamental frequency of vibrations. A controlled experiment using a monopole antenna driven at 900 MHz and vibrating at 150 Hz has significant RF spectral content extending to 5 kHz.\\
Although many studies have been performed in double loop antennas and in antennas with different geometrical shape, non-exhaustive examples can be found in \cite{article}, \cite{kraus2002antennas} and citations therein, a systematic study on the effects of the twist angle in two simple geometric cases, as reported in this paper, has never been performed.\\
Our paper introduces a simple method to increase the performance of an antenna in terms of Return Loss (RL) and corresponding bandwidth. The simplicity of the method is a further benefit.\\
Beside the novelty of the theoretically predicted improvement induced by the adequate twist angles, the experimental results indicate a still higher performance.

\section{Antenna structure and design}
The effect of twisting was studied in two antennas. In the not-twisted case, they both are formed by two circular loops, mutually perpendicular, with a common diameter of 32 cm along the vertical direction, corresponding to the z axis. One loop (loop1) lies in the XZ plane, the other (loop 2) in the YZ plane.\\
The loops were made of harmonic steel (0.5 mm in diameter) in order to have an elastic structure. A length of 100 cm was chosen for each loop, corresponding to a frequency of about 300 MHz (entire wavelength antenna). Furthermore, the loops were shielded from electrical contact by heat-shrinkable sheath and by a small difference (few millimeters) in length.\\
Each antenna was placed on an insulating wood platform keeping the metal ends of the loops fixed, thus preventing their mutual movement. On the opposite side, the two loops crossed a mylar cube with 3 cm size, which kept the two perpendicular wires 2 mm apart each other, thus avoiding any electric contact between them. This cube was free to rotate around the Z axis.
Its rotation angle $\alpha$ was assumed as the measure of the twist angle of the antenna, $\alpha$ = 0 deg corresponding to the not-twisted case.\\
No other electrical or electronic devices (besides N-type connectors for signal supply), have been used in order to guarantee the simplest correspondence between the variations in the antenna characteristics and the torsions applied.

\begin{figure}
\centering
\includegraphics[width=1\columnwidth]{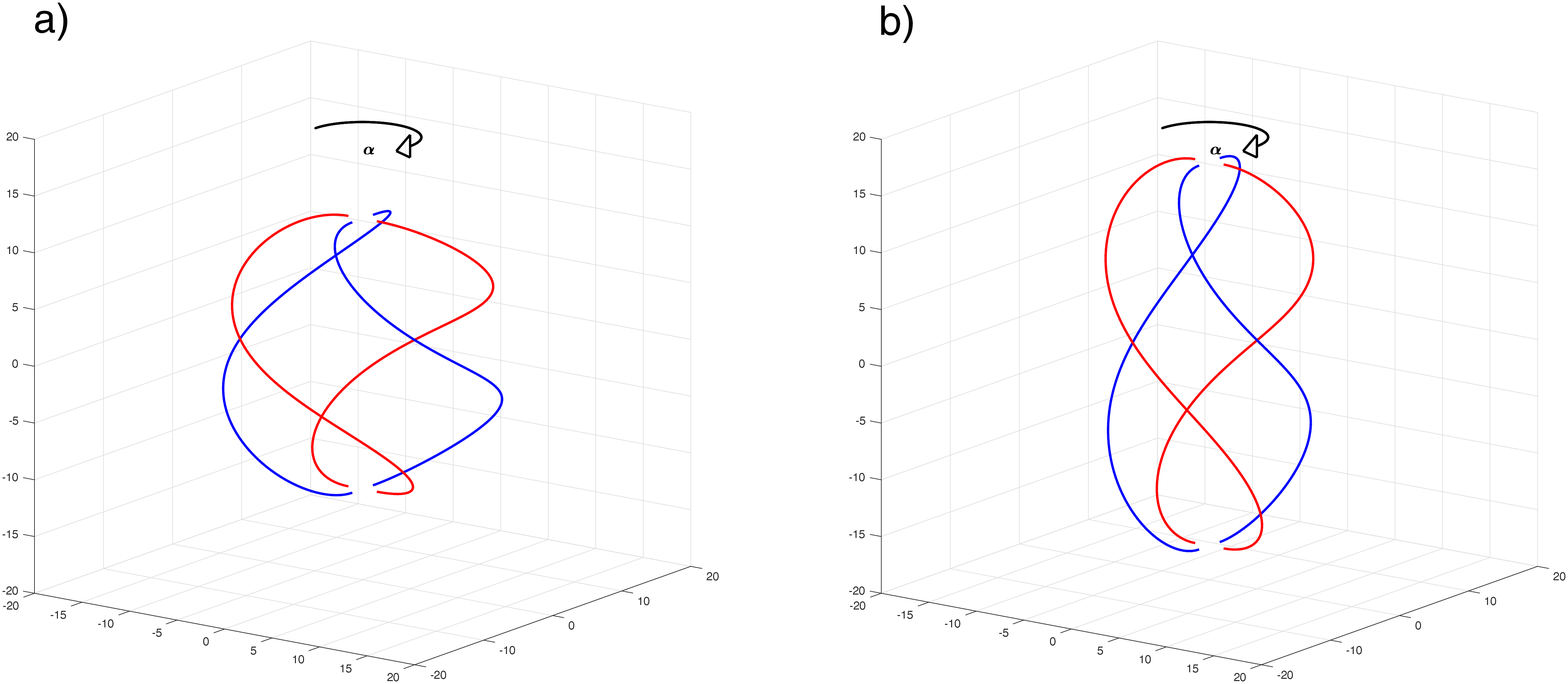}
\caption{Geometry of the two double-loop antennas at 180 deg. of twist angle: Ellipsoidal Geometry (a) and Spherical Geometry(b). In blu: loop1. In red: loop2. (drawn by MATLAB).}
\label{fig:imag2}
\end{figure}
The two antennas, when twisted, corresponded to two different types of geometry.\\
In the first case, the envelope of the twisted antenna keeps the spherical geometry by varying its height (Spherical Geometry - fig. 1a).\\
\begin{figure*}
\centering
\includegraphics[width=0.9\textwidth]{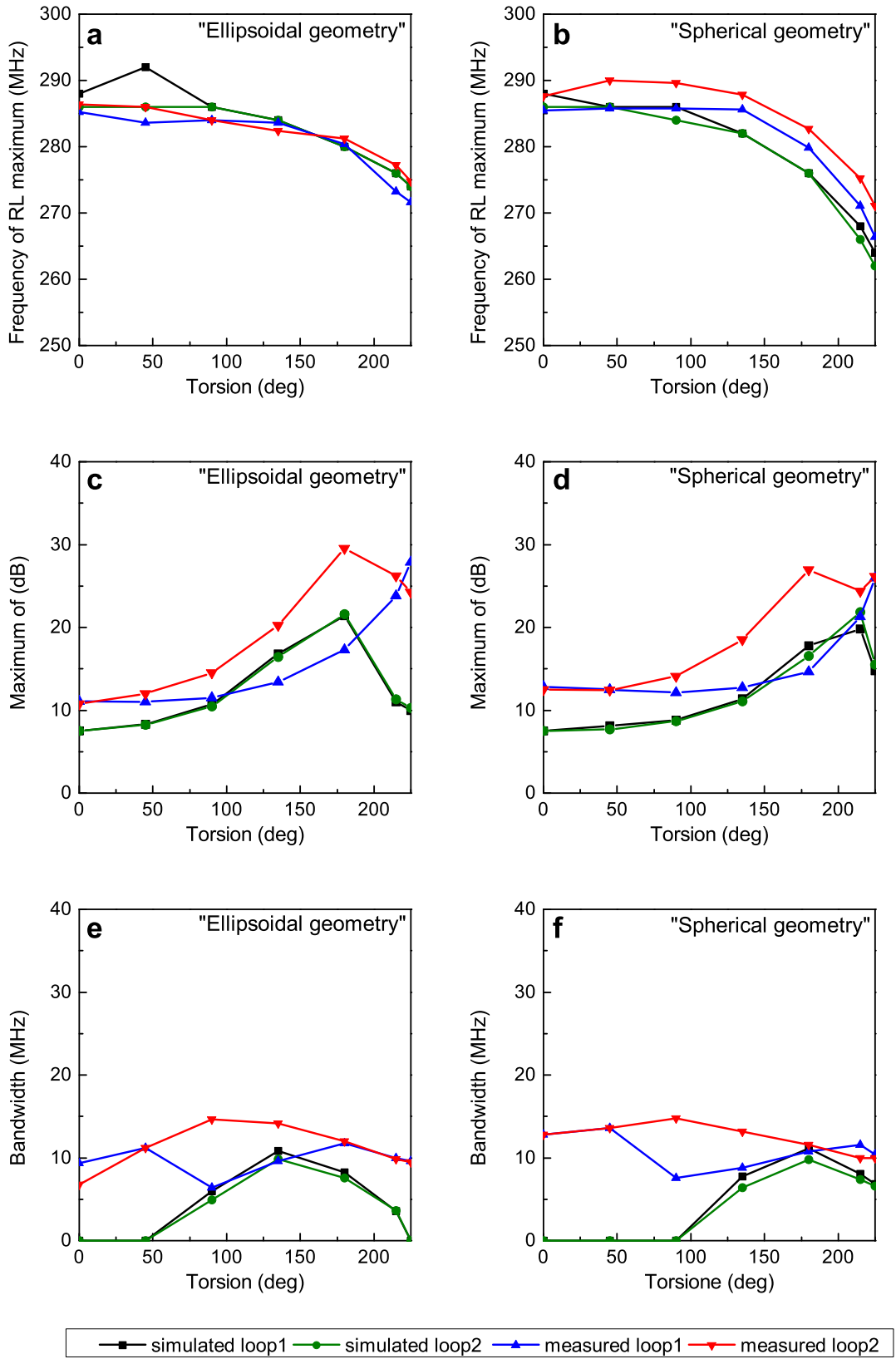}
\caption{Values of the parameters from simulation and from experiment. (a,b) Frequency corresponding to the maximum of the Return Loss, (c,d) Return Loss, (e,f) bandwidth. On the left (a,c,e): "Ellipsoidal geometry" geometry. On the right (b,d,f): "Spherical geometry".}
\label{fig:imag2}
\end{figure*}
In the second case, the antenna was twisted around its symmetry axis z, having fixed height of 32 cm. This value corresponds to the diameter of the two untwisted circular loops. At different twist angles, the torsion causes an ellipsoidal or an hourglass shaped deformation of the antenna envelope (Ellipsoidal Geometry - fig. 1b).\\
A three-dimensional model of the antennas was created to carry out simulations, to predict the expected behavior of the investigated quantities.\\
Each wire of the antennas was modelled as a 3D Kirchhoff rod subjected to twist and bending. Boundary conditions were imposed cinematically at both ends: the twist angle and the slide along the axis of the antenna has been therefore defined as parameters for boundary conditions. The final shape of the wires has been computed in MATLAB, by implementing the numerical method proposed in  \cite{doi:10.1142/9789812774569_0013}.\\
A point cloud of the X, Y, Z coordinates was thus obtained, describing the shape of the two loops of each antenna (fig. 1) A Python script was then developed inside Rhinoceros: starting from the point cloud, it provided us with a STEP file ready for the simulation with the ANSYS HFSS software.\\

\section{MEASURED AND SIMULATED RESULTS}

A HP8510C Network Analyzer \cite{4118975}
was used to obtain the reﬂection coefficients $S_{11}$ and $S_{22}$ of the two loops. They are defined as (Eq.1):

\begin{equation}
RL_i=\left | S_{ii} \right |
\end{equation}

Where $A_i$ and $B_i$ are the amplitudes of the input signal and of the reﬂected signal, respectively, of the ith loop when the
other loop is not power supplied.\\
Also, the Return Loss (RL) is defined as (Eq. 2):

\begin{equation}
S_{ii} = \frac{B_1}{A_1}(i=1,2)
\end{equation}

The Return Loss of the measured and simulated quantities were investigated in the two antennas: frequency corresponding to the maximum of Return Loss (fig.2a and 2b) and maximum of Return Loss (fig.2c and 2d).
The width of the band at RL=10 dB is reported in fig.2e and 2f.\\
In both the antennas, the maximum of RL was observed at a frequency decreasing with increasing twist angle $\alpha$. (fig. 2a and 2b). These variations, were more pronounced in the spherical geometry (about 25 MHz – fig.1b) than in the ellipsoidal one (about 15 MHz – fig 1a), when the twist angle varied between 0 deg. and 225 deg.\\
As far as the Return Loss is concerned, the measured variation as a function of the twist angle is different in the two loops, both in the ellipsoidal geometry and in the spherical one (fig. 2c and 2d) blue curves and red curves, probably due to some difference in welding of the components.\\
In both configurations, the variations observed in loop2 red curves are similar to those predicted by the simulation (green and black curves), although at a shifted absolute value. Thus, in any case a RL higher than predicted is measured.

\section{DISCUSSION AND RESULTS}
In both the antennas, the maximum of RL was observed at a frequency decreasing at increasing twist angle $\alpha$. (fig. 2a and 2b). These variations are more pronounced in the spherical geometry (about 25 MHz – fig.1b) than in the ellipsoidal one (about 15 MHz – fig 1a), when the twist angle varied between 0 deg. and 225 deg. This behavior is in good agreement with that expected by the simulation.\\
The maximum value of the Return Loss as a function of the twist angle shows a different behavior in the two loops, both in the ellipsoidal geometry and in the spherical one (fig. 2c and 2d). Besides other explanations, as for instance an anisotropy in the electromagnetic interaction as predicted by DST theory \cite{cardone2007deformed}, the most simple hypothesis is to assume some difference in the welding of the components. In both the configurations, the variations observed in loop2 are similar to those predicted by the simulation, although at a shifted absolute value. In fact, in any case a RL higher than predicted is measured. Thus, the next step of this investigation will be to check whether they correspond to a higher values of integrated energy emitted by the two antennas.\\
In the case of loop1, a different behavior is observed which, however, is similar in the two geometric configurations. At the highest twist angles (215 deg and 225 deg) the return loss of the two loops is higher than predicted by simulation in both the geometric configurations.\\
finally, the bandwidths measured at $RL_i$ = 10dB is larger than expected by simulation in any case of loop2 , while for loop1 it is larger in the most of the cases, in particular at low twist angles and at high twist angles. The values of Return Loss and bandwith measured experimentally are in general larger than the theoretically expected in both the antennas and both the loops.

\section{Conclusions}
Our paper introduces a simple method to increase the performance of an antenna in terms of Return Loss (RL) and corresponding bandwidth. To this end, the effects of twisting a double loop antenna around its symmetry axis were investigated.\\
Two geometric configurations were considered: one at fixed height and varying shape, the other at fixed spherical shape and varying height.\\
The experimental results were compared with the predictions of the simulation performed by the ANSYS HFSS software. The simulation predicted an improved performance at higher values of the twist angle in terms of Return Loss and band-width in both the geometric configurations.\\
The experimental results, on their side, put in evidence a still better performance both in terms of Return Loss and bandwidth in both the cases.\\
A shift towards lower frequency of the RL maximum for increasing twist angle was also experimentally detected in all the cases, in agreement with the simulation predictions.

\bibliographystyle{IEEEtran}  
\bibliography{mybib}  

\end{document}